\documentclass[prd,twocolumn,showpacs,floatfix,nofootinbib]{revtex4}
\usepackage{graphicx,color,dcolumn,booktabs}
\usepackage{comment,braket}
\usepackage{amssymb,amsmath,bm}
\usepackage{txfonts,float,eufrak}
\usepackage{wrapfig}

\usepackage[colorlinks, citecolor=blue,anchorcolor=red,menucolor=red, linkcolor=red,filecolor=red,runcolor=red,urlcolor=blue,frenchlinks=red]{hyperref}

\def\be{\begin{equation}}
\def\ee{\end{equation}}
\def\bt{\begin{table}}
\def\et{\end{table}}
\def\btb{\begin{tabular}}
\def\etb{\end{tabular}}
\def\bdm{\begin{displaymath}}
\def\edm{\end{displaymath}\mbox{}\\}
\newcommand{\barr}{\begin{array}}
\newcommand{\earr}{\end{array}}
\newcommand{\upr}{\uparrow}
\newcommand{\dar}{\downarrow}
\newcommand{\trip}{$^3P_1$}
\newcommand{\sing}{$^1P_1$}
\newcommand{\qq}{$q\bar{q}$}

\newcommand{\ddas}{$DD^*$}
\newcommand{\dasdas}{$D^*D^*$}
\newcommand{\bbas}{$BB^*$}
\newcommand{\basbas}{$B^*B^*$}
\newcommand{\sw}{$S$-wave}
\newcommand{\dw}{$D$-wave}
\newcommand{\ro}{$r_0$}
\newcommand{\onepp}{1^{++}}
\newcommand{\onepm}{1^{+-}}

\begin{document}

\title{Radially excited axial mesons and the enigmatic $Z_c$ and $Z_b$ in a coupled-channel model}

\author{Susana Coito}\email{susana@impcas.ac.cn}
\affiliation{Institute of Modern Physics, Chinese Academy of Sciences, Lanzhou 730000, China}

\begin{abstract}
The enigmatic charged states $Z_c(3900)$, $Z_c(4020)$, $Z_c(4050)$, $Z_b(10610)$, and $Z_b(10650)$ are studied within a coupled-channel Schr\"odinger model, where radially excited quark-antiquark pairs, with the same angular momenta and isospin as the $a_1(1260)$ and $b_1(1235)$, are strongly coupled to their Okubo-Zweig-Iizuka - allowed decay channels $D\bar{D}^*+\bar{D}D^*$ and $D^*\bar{D}^*$, or $B\bar{B}^*+\bar{B}B^*$ and $B^*\bar{B}^*$, in $S$ and \dw. Poles, matching the experimental mass and width of the above states,  are found by varying only two free parameters. From the wave-function analysis of each resonance, the probability of each of the components contributing to the coupled system is estimated, and predictions can be made for the relative decay fractions among the coupled open-charm or open-bottom decay channels.  
\end{abstract}

\pacs{14.40.Rt, 12.39.Pn, 11.80.Gw}

\maketitle

\section{\label{sec1}Introduction}

{A new family of mesonic resonances with electric charge has been detected by Belle and BESIII Collaborations in the charmonium and bottomonium energy regions. Due to their similarity in mass and in the decay channels, three sets of resonances are considered to form isospin triplets, the $Z_b(10610)^{\pm/0}$ \cite{prl108p122001}, the $Z_c(3900)^{\pm/0}$ \cite{prl110p252001,prl112p022001}, and the $Z_c(4020)^{\pm/0}$/$Z(4025)^{\pm/0}$ \cite{prl111p242001}. A $Z_b(10650)^\pm$ signal was found simultaneously with the $Z_b(10610)^\pm$. All these enhancements were observed in hadronic decay channels, the $Z_c(3900)$ and the $Z_b$ with favored quantum numbers $J^P=1^+$ \cite{pdg}. A broader signal labeled $Z_c(4050)^\pm$ has also been reported \cite{prd78p072004}.} Due to their electric charge, none of the {\color{red}$Z$} resonances can be a pure $c\bar{c}$ or $b\bar{b}$ state, which is the reason why they are often called {\it exotic} heavy mesons.  Another special feature of the $Z$ family is that, with the exception of the $Z_c(4050)$, all resonances lie very close, yet seemingly {\it above}, some threshold. The $Z_c(3900)$ and $Z_c(4020)$ lie near the \ddas\ and \dasdas\ thresholds, and the $Z_b(10610)$ and $Z_b(10650)$ lie near the \bbas\ and \basbas\ thresholds, respectively, where we define $MM^*:=(M\bar{M}^*+\bar{M}M^*)/\sqrt{2}$ and $M^*M^*:=M^*\bar{M}^*$, $M=D$ or $B$. Moreover, two more $Z$ family members were found with higher masses, namely the $Z_c(4250)^\pm$ and the $Z_c(4430)^\pm$ \cite{pdg}, which are also lying very close to the $DD_1$ and $D^*D_1$ thresholds. This very important fact is an evidence for threshold effects to play a crucial role in whatever mechanism might be behind the generation of the $Z$ family, and it should be kept in mind in any theoretical approach. The unveiling of the enigmatic structure of such {\it exotics} will give us new insights on the complex nonperturbative phenomena dominating the strong interactions at intermediate energies.

Attempts have been made to describe the $Z_c$ and $Z_b$ states as tetraquarks or molecules, since the latter have been expected for long time. Results, when favorable, are very imprecise and not decisive for the establishment of the true degrees of freedom of the considered resonances. For tetraquarks the predicted masses are too far below thresholds, with very large errors, of the order of 100 to 300 MeV, and width estimates are nonexistent \cite{Zbtetra,Zctetra}. In Ref.\ \cite{prl111p162003} the author makes the deceiving affirmation that the $Z_c(3900)$ is a tetraquark, while using this very resonance as an input to extrapolate a hypothetical tetraquark spectrum, without width predictions. Molecular interpretations of the $Z$ family are a natural supposition, yet several model results do not favor such scenarios. In molecular models, usually one or more mesons play the role of mediators between the two open-charmed/bottom mesons that form the molecule. In Refs.\ \cite{prd90p016003} and \cite{prd92p034004} it is shown that a slightly bound state, that can be identifyed with the $Z_c(3900)$, may be formed by the influence of $J/\psi$ exchange between $D$ and $D^*$. The argument is sustained by the hypothesis that the {\it true} $Z_c(3900)$ pole would lie {\it below} the \ddas\ threshold, meaning that the signal observed in this channel would be the tail of the resonance structure, while the position of the peak seen in the Okubo-Zweig-Iizuka (OZI)-suppressed decay channels would be shifted from the pole. In fact, the energy of a state does not necessarily coincide with the top of a resonance peak. This point is delicate since it implies that the actual mass and width parameters of the $Z$ family are being read off incorrectly from the experiment. Indeed, the need of more suitable  methods for data analysis has been pointed out in some discussions \cite{1511.06779,ppnp67p449}. In Ref.\ \cite{prd84p054002}, considering the exchange of light mesons and gluons between $B-B^*$ or $B^*-B^*$, the authors can find molecular bound states, but in Ref.\ \cite{prd90p076008} they cannot. The $Z_c(3900)$ molecule is also disfavored by the light front model \cite{epjc73p2561}, and by some lattice QCD results  \cite{prd89p094506,plb727p172,1411.1389}. In some cases, QCD sum rules can produce molecules with very imprecise results \cite{molsrules}. If the $Z$ signals are really {\it above} thresholds, the molecular hypothesis is very unlikely.

{Finally, there is the possibility that the $Z$ signals are due to kinematic effects, with origin in threshold singularities and rescattering of final states. Such effects may still be combined with true poles. Using regularized bubble diagrams, Bugg \cite{epl96p11002} and, more recently, Swanson \cite{prd91p034009}, describe several $Z$ structures as two-body coupled-channel {\it cusp} effects. A different kinematic treatment uses {\it anomalous triangle singularities}, instead of bubbles, leading to more branching points than the simpler two-body case, see Szczepaniak \cite{plb747p410} and Liu, Oka, and Zhao \cite{plb753p297}. In Ref.\ \cite{prd84p094003}, Chen and Liu produce threshold enhancements using an effective lagrangian approach, but without any singularity analysis. In Ref.\ \cite{prd91p051504}, $S$-matrix poles are required for the reproduction of the $Z_c(3900)$ line-shapes, in both $J/\psi\pi$ and $DD^*$ channels. More far-fetched works consider that the $Z_c(3900)$ and $Z_c(4025)$ would be cusp effects generated by molecules composed by $DD_1$ and $D^*D_1$ \cite{prl111p132003}. Kinematic analysis near thresholds are crucial to distinguish true resonances from non-resonant enhancements.} 

One can ask if a quark model approach to the $Z$ family must be completely discarded. For sure, the naive meson description in terms of pure quark-antiquark \qq\ pairs cannot work. Experiment has shown that such simplified picture only works below all thresholds of OZI-allowed decay channels, i.e., channels where two mesons are generated from the breaking of the string between the $q$ and $\bar{q}$, alongside with the creation of a new \qq\  pair from vacuum. Above these thresholds, practically all resonances are nonperturbatively shifted from any spectrum derived from the naive quark model, and other hadronic degrees of freedom must be incorporated in the description of a resonance, as to {\it unquench} the state. Otherwise, one may draw the conclusion
that practically {\it all} states found above radial energy level $N=2$ should be considered {\it exotics}. Indeed, a clear spectrum of regular mesons cannot be disentangled without the unquenching, because there are no {\it pure} \qq\ mesonic states in nature, besides perhaps the ground states. 

In this spirit, Nikolai Kochelev \footnote{Private Communication} suggested an analogy between the mesonic resonance $Z_c(3900)$ and the meson $X(3872)$. Based on the assumption that the structure of the $X(3872)$ is mainly a $c\bar{c}$ core strongly coupled to open-charmed mesons, one can formulate the hypothesis that the $Z_c(3900)$ may be essentially composed by a radially excited \qq\ core, with $q=u$ or $d$, coupled to the same type of open-charmed mesons. The light-quark core would be an axial isovector. Considering an even higher excitation, the $Z_b$ could be proposed in the same way. At first, the idea may seem unrealistic, because a very high radial excitation has many open decay channels, and therefore, the decay fraction to each channel should be too small to be seen. But it could be that the coupling of a certain radial level to the nearby open-charm and open-bottom meson-meson thresholds is particularly high, in such a way that a coupled system would be formed, containing both quark core and decay channels. In the present work, such coupled system is solved with quantum mechanics {within the scattering theory}. The formalism has been formerly employed to the $X(3872)$ in Ref.\ \cite{epjc73p2351}, showing that the $X(3872)$, alias $\chi_{c2}(2P)$, is not a pure molecule, but instead, a strongly unquenched $c\bar{c}$ state with $J^{PC}=\onepp$. This result was confirmed in Ref.\ \cite{epjc75p26}, {where additional OZI-allowed closed decay channels were included. 

The model is described in Sec.\ \ref{sec2}, with details given in Appendices \ref{AA} and \ref{AB}. Poles and wave-function (WF) results for $Z_c$ resonances are presented in Sec.\ \ref{sec3}, and for $Z_b$ states in Sec.\ \ref{sec4}. Summary and conclusions are given in Sec.\ \ref{sec6}.}

\section{\label{sec2}A coupled-channel Schr\"odinger model}

The formalism employed here was developed in the coordinate-space representation in Ref.\ \cite{zpc19p275}, and has been successfully employed in Refs.\ \cite{epjc73p2351} and \cite{epjc75p26} for the axial vector $X(3872)$. It is a coupled-channel Schr\"odinger model for mesons with two or more wave-function components, with a confining potential between a quark-antiquark \qq\ pair, viz.\ a Harmonic Oscillator (HO) with frequency $\omega=190$ MeV, a value that has been fixed in Ref.\ \cite{prd27p1527} and unchanged in all applications of the same kind of model, both in coordinate and in momentum space. {Other confining potentials can, in principle, be adopted. It is obvious that the simple HO, if nonperturbative effects are neglected, is too naive to accuratly desbribe the meson spectra. However, it is also true that any simple quenched potential, even taking into account spin-orbit corrections, is very inaccurate to describe resonances above thresholds. In particular, the well known Coulomb-plus-linear potential fails dramatically to predict a wide number of mesonic resonances \cite{apppsb5p1007}. Here, we adopt the HO, for a matter of simplicity and extension of the study, but also due to the good results it has produced previously in systems with different quark flavors and angular momenta, e.g. \cite{epjc73p2351,epjc75p26,zpc19p275,prd27p1527,prl91p012003}.} 
 
{
\subsection{\label{sec2a} The $n\ q\bar{q}-m\ MM$ system}

Here, a short description of the coupled-channel model is presented. Some computational details may be found in Appendices \ref{AA} and \ref{AB}. A system is composed by $n$ \qq\ confined components with hamiltonian $h_\alpha$, with $\alpha=1,...,n$, given by
\be
\label{hcg}
h_\alpha=\frac{1}{2\mu_\alpha}\bigg(-\frac{d^ 2}{dr^ 2}+\frac{\ell_\alpha(\ell_\alpha+1)}
{r^2}\bigg)+\frac{\mu_\alpha\omega^2r^2}{2}+m_{q\alpha}+m_{\bar{q}\alpha} \; ,
\ee
coupled to $m$ meson-meson $MM$ final components with hamiltonian $h_j$, with $j=1,...,m$, given by
\be
\label{hfg}
h_j=\frac{1}{2\mu_j}\bigg(-\frac{d^ 2}
{dr^ 2}+\frac{\ell_j(\ell_j+1)}{r^2}\bigg)+m_{M_{1j}}+m_{M_{2j}} \; ,
\ee
and it obeys the stationary Schr\"odinger equation
\be
\label{schrg}
\left(\barr{cc}
h_\alpha & V\tilde{g}_{\alpha j}\\
\lbrack V\tilde{g}_{\alpha j}\rbrack^T & h_{j}
\earr\right)
\left(\barr{c}
u_\alpha\\
u_j
\earr\right)=
E\left(\barr{c}
u_\alpha\\
u_j
\earr\right)\ .
\ee
In Eqs.\ \eqref{hcg} and \eqref{hfg}, $\ell_\alpha$, $\ell_j$, $\mu_\alpha$, $\mu_j$ are the orbital angular momenta and reduced mass of each confinement and final components, respectively, $m_{q,\bar{q}\alpha}$ are the constituent quark masses of the confinement state $\alpha$, and $m_{M_{1,2j}}$ are the masses of the final mesons in channel $j$. In the Schr\"odinger equation \eqref{schrg}, $u_\alpha(r)$ and $u_j(r)$ are related to the radial WF $R(r)$ through $u(r)=rR(r)$, and $E$ is the total energy of the system. The whole problem is considered to be spherically symetric. The three-dimensional HO potential in Eq.\ \eqref{hcg} generates a spectrum given by
\be
\label{hog}
E=\Big(2\nu_\alpha+\ell_\alpha+\frac{3}{2}\Big)\,\omega+m_{q\alpha}+m_{\bar{q}\alpha}.
\ee 
where the radial quantum number is given by $\nu$, and the HO frequency by $\omega$.
In  $h_j$, Eq.\ \eqref{hfg}, both mesons are free, i.e., without any final state interaction. In this way, $M_{1j}$ and $M_{2j}$ are connected exclusively by their coupling to the \qq\ bare channels through an off-diagonal potential $V$ given by
\be
\label{potg}
V=\frac{\lambda}{2r_0}\delta(r-r_0)\ ,\ \ \tilde{g}_{\alpha j}=\frac{g_{\alpha j}}{\mu_\alpha}\ ,
\ee 
where $\lambda$ is the global coupling constant, introduced as a free parameter, \ro\ is a transition radius related to the string-breaking distance of the \qq\ pairs, which is also a free parameter, and $g_{\alpha j}$ are the partial couplings between the confinement channel $\alpha$ and the final channel $j$, that should not be free. Although there is no explicit dependence in time, a temporal relation between both components is implicitly assumed since both mesons $MM$ in the {\it final} state must result from the decay of some {\it initial} \qq\ pair. Yet, since the WF is stationary, both states coexist simultaneously. The physical relevant quantities for our study will be the generated poles and the WF probability distributions in space. The $MM$ center-of-mass momentum $k_j$, to appear in the solution $u_j(r)$, and reduced mass $\mu_j$ are relativistic, and given by
\be
\label{mom}
k_j(E)=\frac{E}{2}\left\lbrace\left[1-\left(\frac{M_{1j}+M_{2j}}{E}\right)^2\right]
\left[1-\left(\frac{M_{1j}-M_{2j}}{E}\right)^2\right]\right\rbrace^{\frac{1}{2}}\ ,
\ee
\be
\label{redm}
\mu_j(E)=
\frac{E}{4}\left[1-\left(\frac{M_{1j}^2-M_{2j}^2}{E^2}\right)^2\right] \ .
\ee}

\subsection{Partial Couplings}

In the present model, all final state mesons are connected exclusively {\it through} the bare \qq\ components, and the confinement components are connected exclusively through the final mesons, but nothing is said about the binding mechanisms. However, all the considered $MM$ channels are OZI-allowed. At first, because the coupling to this type of channels is dominant. Secondly, because we wish to avoid the introduction of extra free parameters through the partial couplings $g$ in Eq.\ \eqref{potg}. Indeed, a model has been developed by van Beveren in Ref.\ \cite{pcoup} to evaluate such couplings for the case of OZI-allowed channels. There, when the string between $q$ and $\bar{q}$ is broken, a new \qq\ pair with quantum numbers $^3P_0$ is created from the vacuum, and it recombines with the initial quarks. The transitions are computed using a HO basis and angular momentum conservation, via Clebsch-Gordon coefficients. For \qq\ pairs with quantum numbers $J^{PC}=\onepp$ and $\onepm$, or \trip\ and \sing, respectively, the coupling to a pseudoscalar (P) - vector (V) and to a VV meson pair with orbital momentum $\ell_j$ is given in Tab.\ \ref{tcoup}, where all mesons P and V are in the ground state. The radially dependent partial couplings $g_n$ in Eq.\ \eqref{potg}, with $n=0,1,2,...$, are given by $g_n=g_0\times c_n$. An extra factor $1/4^n$ over all $g^2$ couplings is predicted in the model \cite{pcoup}. This factor ensures that the larger the $n$, the smaller the coupling of a certain radially excited \qq\ state to each decay channel will be. On the other hand, the number of open decay channels increase with the $n$, in such a way that highly excited systems may still be significantly unquenched. In the present study, we neglect the $1/4^n$ factor, resulting in significantly larger couplings between our \qq\ systems and the decay channels under consideration. These {\it effective} couplings simulate the effect of the many decay channels which are not included in this description. It is assumed that the off-diagonal transition potential in Eq.\ \eqref{potg} should be localized in space. Here it is point-like, mostly for the sake of simplicity.           

\bt
\centering
\btb{c|c|c|c|c}
\hline\hline
&&&&\mbox{}\\[-3mm]
&$\ell_j$&$g^2_{n=0}(\onepp)$&$g^2_{n=0}(\onepm)$& $c^2_n$\\[1mm]
\hline
&&&&\mbox{}\\[-3mm]
PV&0&1/18&1/36&$n+1$\\[1mm]
PV&2&5/72&5/36&$2n/5+1$\\[1mm]
VV&0&0   &1/36&$n+1$\\[1mm]
VV&2&5/24&5/36&$2n/5+1$\\[1mm]
\hline\hline
\etb
\caption{\label{tcoup}Partial couplings, computed from the model in Ref.\ \cite{pcoup}. The total coupling $g_n$, with $n=0,1,..$, is given by $g_n=g_0\times c_n$. P and V are non-excited pseudoscalar and vector mesons, respectively.}
\et

\section{\label{sec3}The $Z_c(3900)$, $Z_c(4020)$, and $Z_c(4050)$}

{
As mentioned in Sec.\ \ref{sec1}, the $Z_c(3900)$ and $Z_c(4020)$ were assigned the quantum numbers isospin $I=1$, and favored $J^P=1^+$ in the experiment. Within the present model the isospin is accounted for by the \qq\ component, where $q=u$ or $d$, without distinguishing among isospin triplet states. Since we do not know the $C$-parity of the $Z_c$ resonances, we admit that they could be mixtures of singlet \sing\ and triplet \trip\ states, that within this model couple to the same decay channels, but we also consider the states separately. For the final states $j$ we use the \ddas\ and \dasdas\ channels both in \sw\ and \dw. The constituent quark mass is {\color{red}$m_q=406$} MeV \cite{prd27p1527}, and the meson masses are taken from the experiment \cite{pdg}. Since $\omega$ is considered as an {\it universal} constant, the only actual free parameters are $\lambda$ and \ro. Both of them are tuned as to reproduce the approximate experimental mass and width of the $Z_c(3900)$, viz.\ $3889-i17$ MeV \cite{pdg}. As a pole {\it above} threshold, it acquires an imaginary part as the coupling $\lambda$ is turned on, describing a parabolic-like trajectory, cf.\ Eq.\ \eqref{det}. The free parameters are very restricted if we are to reproduce the $Z_c(3900)$ in mass and width, i.e., $0.63<r_0<0.69$ fm, and $1.0<\lambda<1.2$. Within this range, and with two channels only, i.e., $q\bar{q}\ (1^+)-DD^*\ (\ell=0)$ or the whole set of five and six coupled-channels, respectively  $q\bar{q}\ (1^{++}\ or\ 1^{+-})-(DD^*+D^*D^*)\ (\ell=0,2)$ and $q\bar{q}\ (1^{++}+1^{+-})-(DD^*+D^*D^*)\ (\ell=0,2)$, we can always find a pole around $3889-i17$ MeV coming from the {\it confinement} (conf) spectrum. We also find a {\it dynamical} (dyn) pole from the continuum, with a very large width. Using the same parameters, no poles are found corresponding to the $Z_c(4020)$ and $Z_c(4050)$.

In Fig.\ \ref{f9} we show the phase shift dependence with energy when $r_0=0.65$ fm and $\lambda=1.1$, for the system  $q\bar{q}\ (1^+)-DD^*\ (\ell=0)$. It can be seen the $-90^o$ phase shift around 3.89 and 4.06 GeV, corresponding to the energy  of resonance poles. The width of each slope corresponds to the width of each resonance. This typical phase shift behavior shows consistency within the model.   

\begin{figure}
\resizebox{!}{190pt}{\includegraphics{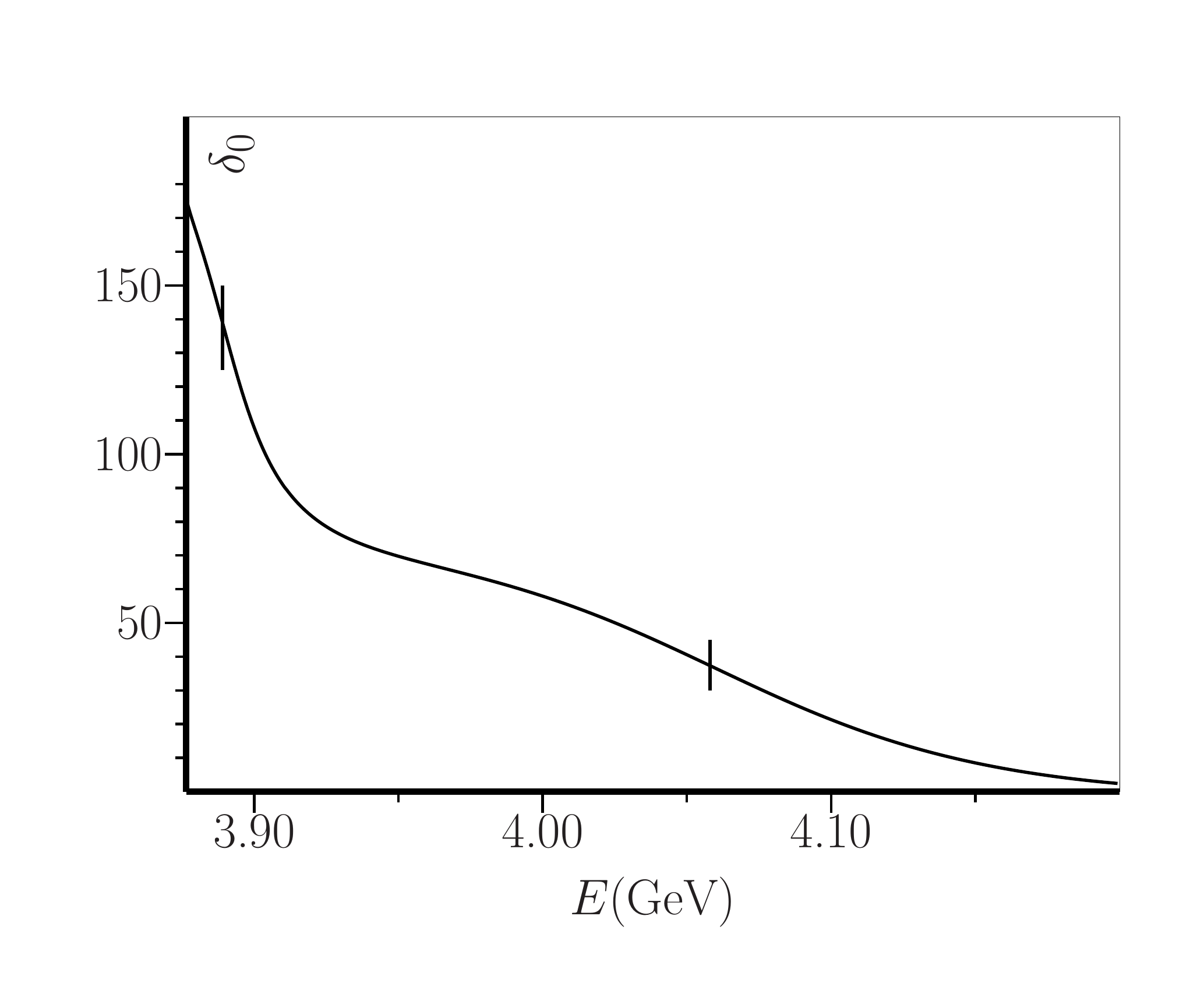}}
\caption{\label{f9} Phase Shift, in degrees, for the system $q\bar{q}\ (1^+)-DD^*\ (S-wave)$, with $r_0=0.65$ fm and $\lambda=1.1$. Vertical lines correspond to the real energy of the confinement pole at $3889-i17$ MeV, and the dynamical pole at $4058-i103$ MeV.}
\end{figure}

If we consider the mixture $^1P_1 +\ ^3P_1$ for the confining component we find a new scenario, with higher values for the free parameters, where the three $Z_c(3900)$, $Z_c(4020)$ an $Z_c(4050)$ are produced simultaneously, with $Z_c(4020)$ an $Z_c(4050)$ being generated dynamically. Poles and WF probabilities are described in Tab.\ \ref{t5}. If the \sing\ and \trip\ states do not mix, the $Z_c(4020)$ and $Z_c(4050)$ cannot be reproduced simultaneously with the $Z_c(3900)$. The WF probabilities in Tab.\ \ref{t5} allow us to estimate the relative decay fractions among the considered channels, e.g., both poles around 4.02 and 4.05 GeV should be seen in both \ddas\ and \dasdas\ channels in \sw. However, unlike the 3.89 GeV pole, the energy of both these dynamical poles is quite sensitive to the values of the free parameters. In the same way, the mass of the dynamical poles might, as well, be sensitive to other unquenching effects, e.g., the proximity of the axial-vector $X(3872)$, or the coupling to all the other OZI-allowed decay channels, open and closed. Also, the hypothesis that the $Z_c(4020)$ and $Z_c(4050)$ might not be axials should not be excluded. In case of the $Z_c(4050)$, the quark flavor content might even be different, as discussed in Refs.\ \cite{posp003,plb669p156}, where this resonance is considered a radial excitation of the $c\bar{s}$ system.

\bt
\centering
\vspace*{2mm}
\hspace*{-3mm}
\btb{c|c|ccccccc}
\hline\hline&&&&&&&\\[-3mm]
Poles&Type&$P(R_{\onepp})$&$P(R_{\onepm})$&$P(R_{DD^*}^{\ell=0})$&$P(R_{DD^*}^{\ell=2})$&$P(R_{D^*D^*}^{\ell=0})$&$P(R_{D^*D^*}^{\ell=2})$\\[1mm]
\hline&&&&&&&\\[-3mm]
$3890-i3$ &conf&25.9&10.8&61.6&1.5&0.0&0.2\\[1mm]
$4006-i28$&conf&46.4&40.9&9.8&0.3&1.2&1.4\\[1mm]
$4027-i7$ &dyn&10.6&37.2&27.3&0.4&22.4&2.2\\[1mm]
$4053-i20$&dyn&25.4&39.5&13.8&0.8&13.5&7.1\\[1mm]
\hline\hline
\etb
\caption{\label{t5} Poles found for the system $q\bar{q}\ (^1P_1 +\ ^3P_1) - (DD^*+D^*D^*)\  (S+D-wave)$. $r_0=0.89$ fm and $\lambda=6.3$. Wave-function probabilities for all components, in \%.}
\et  

\begin{figure}
\centering
\resizebox{!}{360pt}{\includegraphics{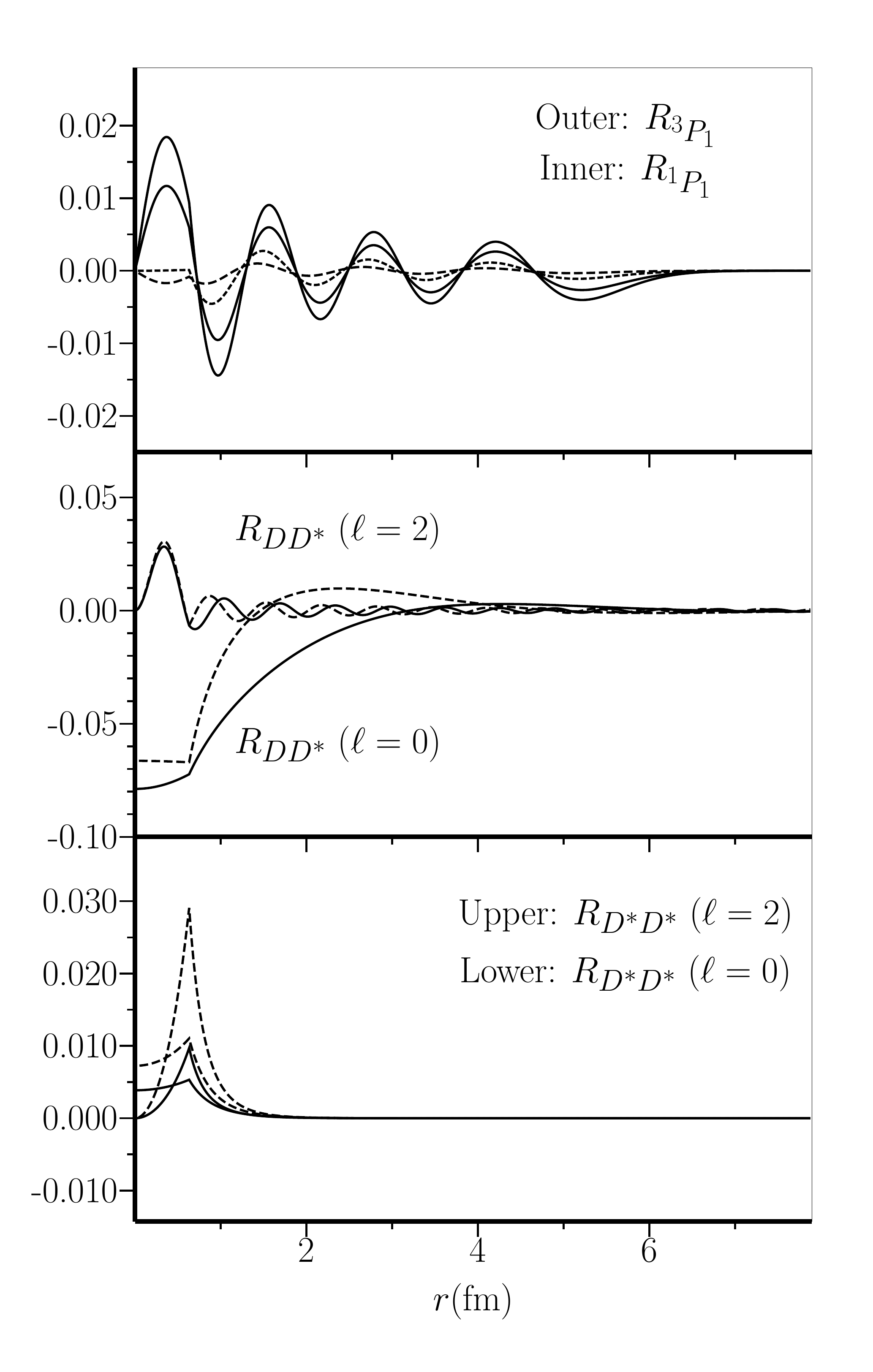}}
\caption{\label{f5} Radial wave-function, in GeV$^{1/2}$, for the pole $3889-i14$ MeV, with \ro=0.63 fm and $\lambda=1.0$, in the six coupled-channel system. Solid lines: $\mathfrak{Re}\ (R(r))$. Dash lines: $\mathfrak{Im}\ (R(r))$. For meson-meson components, \dw\ curves start in zero, while \sw\ curves start in a non-zero value. For \qq\ components, the real and imaginary waves with greater amplitude are the \trip.}
\end{figure}

The WF corresponding to the confinement pole $3889-i14$ MeV, for \ro=0.63 fm and $\lambda=1.0$, is depicted in Fig.\ \ref{f5}. The confinement components exhibit a nodal structure, as expected for a high radial quantum number $n_r$, the \ddas\ \sw\ component converge to zero with a slight oscillation, as it corresponds to a resonance solution, the \ddas\ \dw\ is much more nodal than the \sw\ and it scatters over a larger space before convergence. For the \dasdas\ component, both $S$ and \dw\ converge to zero without nodes, since this channel is closed.}

\section{\label{sec4}The $Z_b(10610)$ and $Z_b(10650)$}
\begin{figure}
\resizebox{!}{190pt}{\includegraphics{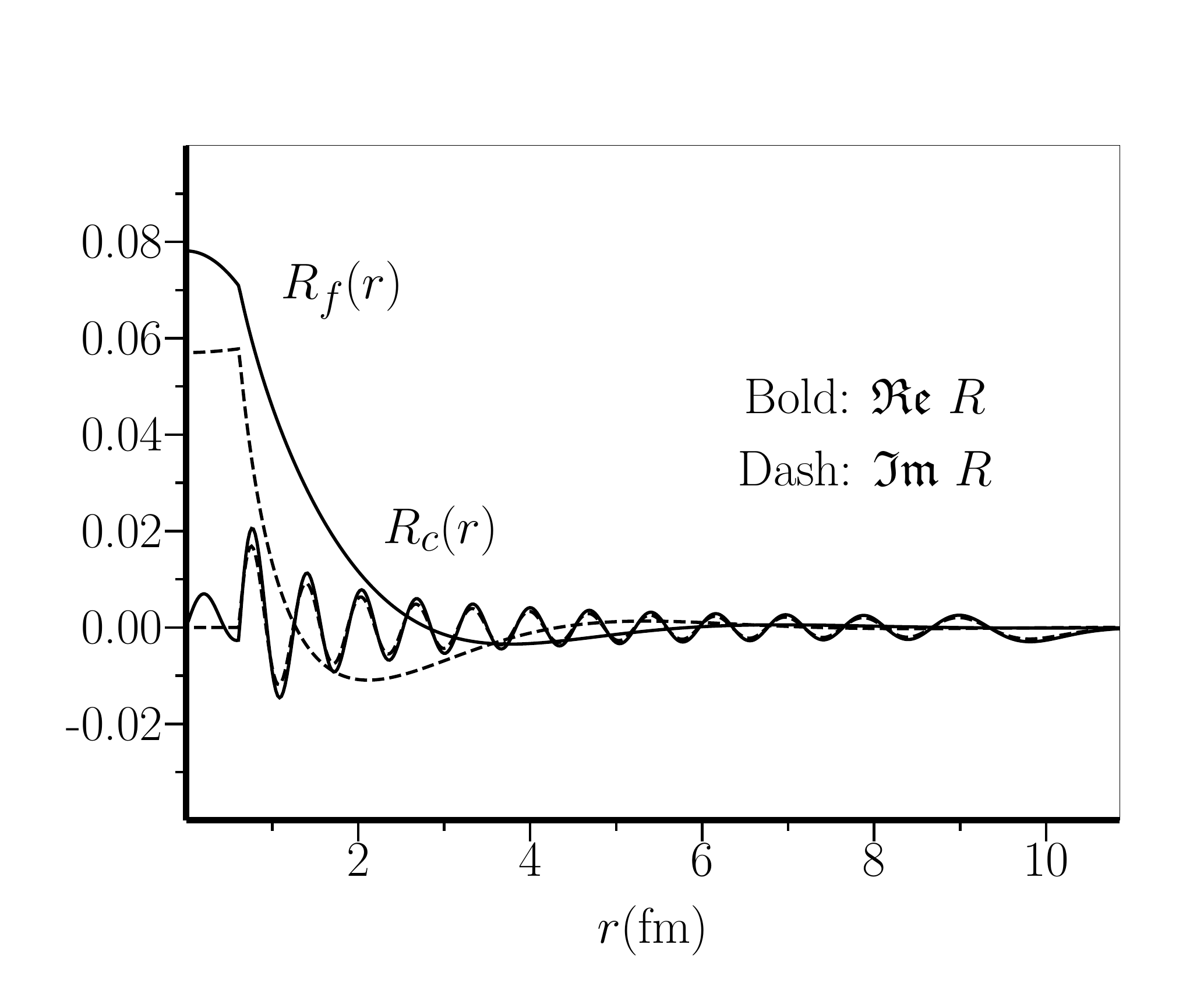}}
\caption{\label{f7}Radial wave-function $R(r)$, in GeV$^{1/2}$, for the pole $10611-i6$ MeV found for the system $q\bar{q}\ (1^+)-BB^*\ (S-wave)$, with \ro=0.61 fm and $\lambda=2.2$. Solid line: $\mathfrak{Re}\ (R_{c,f})$. Dash lines: $\mathfrak{Im}\ (R_{c,f})$. Curves with nodes: $R_c(r)$, where $c$ is the confinement \qq\ channel. Other curves: $R_{f}(r)$, where $f$ is the \bbas channel in \sw.}
\end{figure}

{
The same analysis is performed as in Sec.\ \ref{sec3}, but for the energy region of the bottomonium. The coupled system involves two excited \qq\ states with $q=u$ or $d$ and $J^{PC}=\onepp$ and $\onepm$, either mixed or unmixed, and the decay channels \bbas\ and \basbas\, in $S$ and \dw. We take as reference the experimental mass and width for the $Z_b(10610)$ around $10607-i9$ MeV \cite{pdg}, and tune both free parameters as to reproduce this state. Indeed, a narrow pole is found near and above the \bbas\ threshold for two different radius \ro, but of dynamical type. Using the same free parameters, such pole is found either for the simplest system with two coupled-channels only, or for the system with six or five coupled-channels, with either the singlet and triplet \qq\ states mixed or unmixed, correspondingly. A narrow dynamical pole around 10.65 GeV also arises as an effect of coupling the \basbas\  channels. If the spin singlet and triplet states do not mix, the latter pole is only generated for the higher radius $r_0=0.93$ fm, for both singlet and triplet states. If they do mix, a 10.65 GeV pole is produced for both radius. The idea of mixed spin in both $Z_b$ resonances has been discussed in Ref.\ \cite{prd84p054010}, although the confined states were considered to be of bottomonium type. Results with mixed confinement states are shown in Tab.\ \ref{t9}. The probability of the decay channels may provide a way of distinguishing scenarios. Most poles from the confinement decouple from the decay channels and are not expected to be seen experimentally. Such poles appear as the closer systems to molecules, i.e., two \qq\ states connected through intermediate $B$ and $B^*$ mesons.\\ 
The WF corresponding to the pole at $10611-i6$ MeV, for the two coupled-channel system $q\bar{q}\ (1^+)-B^*B^*$ (\sw), is plotted in Fig.\ \ref{f7}. As a higher radial excitation, the confinement component of the WF becomes more nodal than in case of the 3.89 GeV, cf.\ Fig.\ \ref{f5}.}

\bt
\hspace*{-5mm}
\centering
\btb{c|c|cccccc}
\hline\hline&&&&&&&\\[-3mm]
Poles (MeV)&Type&$P(R_{\onepp})$&$P(R_{\onepm})$&$P(R_{BB^*}^{\ell=0})$&$P(R_{BB^*}^{\ell=2})$&$P(R_{B^*B^*}^{\ell=0})$&$P(R_{B^*B^*}^{\ell=2})$\\[1mm]
\hline&&&&&&&\\[-3mm]
{$10608-i2$}&dyn&14.5&5.5&72.4&7.2&0.1&0.3\\[1mm]
{$10647-i2$}&dyn&14.9&9.4&69.8&4.4&1.1&0.4\\[1mm]
$10711-i47$&conf&47.3&45.1&3.6&0.8&0.9&2.4\\[1mm]
$10817-i7$&conf&55.3&38.6&0.3&0.6&0.1&5.2\\[1mm]
\hline\hline
\etb\\
\vspace*{4mm}
\hspace*{-5mm}
\btb{c|c|cccccc}
\hline\hline&&&&&&&\\[-3mm]
Poles (MeV)&Type&$P(R_{\onepp})$&$P(R_{\onepm})$&$P(R_{BB^*}^{\ell=0})$&$P(R_{BB^*}^{\ell=2})$&$P(R_{B^*B^*}^{\ell=0})$&$P(R_{B^*B^*}^{\ell=2})$\\[1mm]
\hline&&&&&&&\\[-3mm]
{$10607-i4$}&dyn&12.6&4.6&81.5&0.1&0.20&1.1\\[1mm]
{$10650-i3$}&dyn&16.4&17.2&54.7&0.1&9.1&2.6\\[1mm]
{$10665-i9$}&dyn&9.0&13.4&61.5&0.1&9.7&6.3\\[1mm]
$10754-i5$&conf&48.9&32.9&3.1&0.1&0.1&14.9\\[1mm]
$10911-i7$&conf&5.3&6.8&20.8&15.9&16.6&34.6\\[1mm]
\hline\hline
\etb
\caption{\label{t9}Poles found for the system $q\bar{q}\ (^1P_1 +\ ^3P_1) - (BB^*+B^*B^*)\  (S+D-wave)$. Upper table: $r_0=0.61$ fm and $\lambda=2.0$. Lower table: $r_0=0.93$ fm and $\lambda=2.0$. Radial wave-function probabilities for all components, in \%. }
\et

\section{\label{sec6}Summary and Conclusions}
{
A quark-model inspired coupled-channel system has been analyzed by employing quantum mechanics scattering formalism. The aim was to represent the newly discovered enigmatic heavy mesons with charge $Z_c(3900)$, $Z_c(4020)$, $Z_c(4050)$, $Z_b(10610)$, and $Z_b(10650)$. We have coupled a high radial excitation of two \qq\ pairs, with $q=u$ or $d$, and quantum numbers \sing\ and \trip, to the OZI-allowed decay channels \ddas\ and \dasdas, or \bbas\ and \basbas, in $S$ and \dw. By tuning the only two free parameters of the model, the global coupling $\lambda$ and a transition radius \ro, poles have been found matching the experimental mass and width of the above $Z_c$ and $Z_b$ resonances. Namely, we find a stable pole around 3.89 GeV, and more sensitive dynamical poles at 10.61 and 10.65 GeV that are produced simultaneously. Poles at 4.02 and 4.05 GeV might have a different origin, or else result from the mixing between \trip\ and \sing\ states. It is possible that they actually have other angular momenta or, in case of the $Z_c(4050)$, other quark flavor content. One important test is the detection of a neutral partner for the $Z_c(4050)^\pm$. The studied $Z$ states can be regarded as high radial excitations of the $a_1(1260)$ and $b_1(1235)$, that can mix or simply superpose.
The employed model is simplistic and intends to show the feasibility of addressing new enigmatic states as higher radial excitations within the quark model, and to emphasize the relevance of the unquenching. The used HO confining potential can be replaced by other potentials, yet no phenomenological potential is known that has proved to be more reliable above thresholds. 
We point out that the unquenching could generate similar states near $DD_1$, $D^*D_1$, $BB_1$, and $B^*B_1$ thresholds. In fact, two $Z_c$ states have been seen near the former two thresholds, one of them, the $Z_c(4430)$, is an axial state. Also, charged mesonic resonances should be seen with other angular momenta, namely vectors near the $DD$ and $BB$ thresholds. Although no charged vectors have been seen in the experiment, the signal for the neutral $\psi(3770)$ is known to be distorted, and may result from a superposition of states. In the same way, radially excited $u\bar{s}$ and $d\bar{s}$ states should also appear near the open-charm/bottom-strange thresholds. Radial excitations of light quarks far from dominant OZI-allowed decay channels are not expected to be seen, since their nearby decay channels are composed by radially excited mesons with small couplings to the \qq\ core.   

In conclusion, we have shown that if unquenching effects are structurally taken into account, the quark-model does not need to be abandoned when one aims to explain the family of the charged heavy mesons. One also finds that, instead of having a traditional picture of a confined quark-antiquark pair for mesons, new configurations emerge, with quark and meson degrees of freedom, some of which may be of molecular type, but with totally unexpected mass and properties.} 

\section*{Acknowledgments}
{
Thanks to Nikolai Kochelev for the initial idea, to Jean-Marc Richard, Sven Bjarke Gudnason, and Serguei Maydanyuk for useful suggestions, and to Eef van Beveren and George Rupp for important remarks. This project was supported by Chinese Academy of Sciences President's International Fellowship Initiative, Grant No.~2015PM064.}

\appendix

\section{Solving the coupled-channel Schr\"odinger equation}
\label{AA}
{Here, we find the solution of Eq.\ \eqref{schrg}. The solution involves two types of components, with $u_\alpha(r)$ corresponding to the \qq\ confinement components and $u_j(r)$ to the $MM$ free mesons components.

\subsection{\label{AA1} Solutions $u_\alpha(r)$}

From Eq.\ \eqref{schrg} we have
\be
\label{equc}
\begin{split}
&\bigg\lbrace\frac{1}{2\mu_\alpha}\bigg(-\frac{d^ 2}{dr^ 2}+\frac{\ell_\alpha(\ell_\alpha+1)}{r^2}\bigg)+\frac{1}{2}\mu_\alpha\omega^2r^2+m_{q\alpha}+m_{\bar{q}\alpha}-E\bigg\rbrace u_\alpha(r)\\
&=-\frac{\lambda}{2r_0}\delta(r-r_0)\sum_j\tilde{g}_{\alpha j}u_j(r)\ .
\end{split}
\ee}
At $r\ne r_0$ the solution of the above equation is the solution of the homogeneous equation. The delta-shell function will later determine the boundary conditions at $r=r_0$. Considering the homogeneous equation, we perform the following variable change
\be
\label{cvar}
x=\mu \omega r^2\ .
\ee
Then, with the definition
\be
\label{rdf}
u(x)=x^{(\ell+1)/2}e^{-x/2}\phi(x)\ ,
\ee
we get the equation
\be
\label{kumeq}
x\frac{d^2\phi(x)}{dx^2}+(b-x)\frac{d\phi(x)}{dx}-a\phi(x)=0\ ,
\ee
with
\be
a=-\nu,\ \ \ b=\ell+3/2\ ,
\ee
where
\be
\label{niu}
\nu=\frac{E-(m_q+m_{\bar{q}})}{2\omega}-\frac{\ell+3/2}{2}\ .
\ee 
Here, the parameter $\nu$ is equivalent to the radial quantum number of the HO in Eq.\ \eqref{hog}. Equation \eqref{kumeq} is the confluent hypergeometric equation, or Kummer equation. It admits the following solutions
\be
\label{chg1}
\phi(a,b,x)=\sum_{n=0}^{\infty}\frac{(a)_n}{(b)_n}\frac{x^n}{n!},\ (a)_0=1,\ (a)_{n+1}=(a+n)(a)_n\ ,
\ee
or
\bdm
\phi(a,b,x)=1+\frac{a}{b}x+\frac{(a+1)a}{(b+1)b}\frac{x^2}{2!}+\frac{(a+2)(a+1)a}{(b+2)(b+1)b}\frac{x^3}{3!}+\dots\ ,
\edm
and 
\be
\label{chg2}
\begin{split}
\psi(a,b,x)=&\frac{\Gamma(1-b)}{\Gamma(c-b+1)}\phi(a,b,x)+\\&\frac{\Gamma(b-1)}{\Gamma(a)}x^{1-b}\phi(a-b+1,2-b,x)\ ,
\end{split}
\ee
where $\Gamma$ is the complex Gamma function
\be
\Gamma(x)=\int_0^\infty t^{x-1}e^{-t}dt,\ \ \mathfrak{Re}(x)>0\ .
\ee
Equations \eqref{chg1} and \eqref{chg2} are the confluent hypergeometric functions of first and second kind, respectively, the first regular in the origin, while the second falls off in the infinity. 

{
Equation \eqref{niu} represents the radial dependence with the energy. We notice that when the right-hand term of Eq.\ \eqref{equc} is set to zero, the eigenvalues $E$ will correspond to the HO spectrum, and $\nu$ in Eq.\ \eqref{niu} will be integer positive. In such case, the solution would be given in terms of the generalized Laguerre polynomials. When $\lambda\ne0$, and all $MM$ channels are closed, the discrete spectrum is shifted in the real energy axis, and $\nu$ assumes non integer real values.}

With the definitions
\begin{eqnarray}
\label{ffc}
\hspace*{-20pt} F_\alpha(r)=&\displaystyle\:\frac{1}{\Gamma(\ell_\alpha+3/2)}\,x^{(\ell_\alpha+1)/2}\,
e^{-x/2}\,\phi(-\nu,\ell_\alpha+3/2,x) \; , & \\
\displaystyle
\label{fgc}
\hspace*{-20pt} G_\alpha(r)=&\displaystyle-\frac{1}{2\sqrt{\omega\mu_\alpha}}\Gamma(-\nu)\,x^{(\ell_\alpha+1)/2}\,e^{-x/2}
\,\psi(-\nu,\ell_\alpha+3/2,x), &
\end{eqnarray}
where $\Gamma$ functions act simply as convenient constants, the general solution of \eqref{equc} will be
\be
\label{wfc}
u_\alpha(r)=
\left\lbrace\barr{lc}
A_\alpha F_\alpha(r) & ,\ r<r_0 \; \\[5pt]
B_\alpha G_\alpha(r) & ,\ r>r_0 \; 
\earr\right. ,
\ee
where $A_\alpha$ and $B_\alpha$ are constant amplitudes.

\subsection{Solution for $u_j(r)$}

{From Eq.\ \eqref{schrg} it also comes
\be
\label{equf}
\begin{split}
&\bigg\lbrace\frac{1}{2\mu_j}\bigg(-\frac{d^ 2}{dr^ 2}+\frac{\ell_j(\ell_j+1)}{r^2}\bigg)+M_{1j}+M_{2j}-E\bigg\rbrace u_j(r)\\
&=-\frac{\lambda}{2r_0}\delta(r-r_0)\sum_\alpha \tilde{g}_{\alpha j}u_\alpha(r)
\end{split}.
\ee}
As in the previous case, the solution of Eq.\ \eqref{equf} at $r\ne r_0$ will be the solution of the homogeneous equation. This corresponds to the solution of the free wave for any angular momentum $\ell_j$. For $E < M_{1j} + M_{2j}$, it can be shown that the general solution is given by
\be
\label{wff}
u_j(r)=
\left\lbrace\barr{lc}
A_j J_{\ell_f}(kr)& ,\ r<r_0\\\mbox{}\\
B_j\Big\lbrack J_{\ell_j}(kr)k^{2\ell_j+1}\mathrm{cotg}\ \delta_{\ell_j}(E)-N_{\ell_j}(kr)\Big\rbrack  & ,\ r>r_0
\earr\right.\ ,
\ee
with the definitions
\be
\label{bess}
J_\ell(kr)=k^{-\ell}r\ j_\ell(kr),\ N_\ell(kr)=k^{\ell+1}r\ n_\ell(kr)\ ,
\ee
where $k$ is the final $MM$ center of mass momentum, $n_{\ell}(kr)$ and $j_{\ell}(kr)$ are the Neumann and Bessel functions, respectively, and $\delta_{\ell}$ are the phase shifts. In our problem, we wish to describe a resonance, with complex energy above threshold. The wave-function, therefore, must suffer a modification in order to be convergent in the infinity. For {\it resonances} we have $E=(E_1,-E_2)$, with $E_1,E_2 \in \mathfrak{Re}$, and the general solution of \eqref{equf} will be 
\be
\label{wffr}
u_j(r)=
\left\lbrace\barr{lc}
A_j J_{l_j}(kr)& ,\ r<r_0\\\mbox{}\\
B_j\Big\lbrack J_{l_j}(kr)k^{2l_j+1}\mathrm{cotg}\ \delta_{\ell_j}(E)+N_{l_j}(kr)\Big\rbrack  & ,\ r>r_0
\earr\right. .
\ee
The {\it negative} energy solution \eqref{wff}, and {\it complex} energy solution \eqref{wffr}, only exist due to the inhomogeneous term in Eq.\ \eqref{equf} which defines the boundary conditions at $r=r_0$. In the free case there is only solution for $E>M_{1j}+M_{2j}$, and it is simply given by $A_j J_{\ell_j}(kr)$ in all space.\\
{The solutions with complex energy result from the analytic continuation of the real energy to the complex plane, when at least one final channel $j$ is open. This can be regarded in terms of a scattering matrix $\mathcal{S}$ that one can build to represent the problem. Such matrix is unitary for open channels $j$ and can assume complex values. Since $\mathcal{S}$ is meromorphic in the energy, the energy can assume complex values as well.}

\section{Boundary conditions\label{AB}}

Here, we compute the boundary conditions for resonances at $r=r_0$ for the coupled-channel system $n\ q\bar{q} - m\ MM$. For the simplest $q\bar{q} - MM$ case, details may be found in Ref.\ \cite{epjc73p2351}. From Eq.\ \eqref{schrg}, the boundary conditions at $r=r_0$ will be 
\be
\label{bc1g}
\begin{split}
&{u'}_\alpha(r\upr a)-{u'}_\alpha(r\dar r_0)=-\frac{\lambda}{r_0}\sum_jg_{\alpha j} u_j(r_0)\ ,\\
&u_j'(r\upr r_0)-u_j'(r\dar r_0)=-\frac{\lambda}{r_0}\mu_j\sum_\alpha \tilde{g}_{\alpha j} u_\alpha(r_0)\ ;
\end{split}
\ee
\be
\label{bc2g}
\begin{split}
&{u}_\alpha(r\upr r_0)={u}_\alpha(r\dar r_0)\ ,\\
&u_j(r\upr r_0)=u_j(r\dar r_0)\ ,
\end{split}
\ee
with $\alpha=1,..,n$ and $j=1,...,m$. The conditions in Eqs.\ \eqref{bc1g} and \eqref{bc2g} over Eqs.\ \eqref{equc} and \eqref{equf} lead to the amplitude relations
\be
\begin{split}
\label{ampr}
&A_\alpha=\frac{\lambda}{r_0}G_\alpha (r_0)\sum_j g_{\alpha j}J_{\ell_j}(kr_0)\tilde{A}_j\ ,\\
&\tilde{A}_j=\frac{\lambda}{r_0}\mu_jC_{\ell_j}(kr_0)\sum_\nu\tilde{g}_{\alpha j}F_\alpha(r_0)A_\alpha\ ,
\end{split}
\ee
with
\be
C_{\ell_j}(kr_0)=J_{l_f}(kr_0)k^{2l_f+1}\mathrm{cotg}\ \delta_{\ell_j}(E) \pm N_{l_f}(kr_0)\ ,
\ee
where the plus sign holds for open $MM$ channels, and the minus sign for closed $MM$ channels. From Eqs.\ \eqref{ampg} it follows
\be
\label{ampg}
A_\alpha=\bigg(\frac{\lambda}{r_0}\bigg)^2G_\alpha(r_0)\sum_jg_{\alpha j}J_{\ell_j}C_{\ell_j}\mu_j\sum_\beta \tilde{g}_{j\beta}F_\beta(r_0)A_\beta\ .
\ee
Defining the matrices
\be
G=\{G_\alpha\},\ \mathcal{G}=\{g_{\alpha j}\},\ K=\{\mu_jJ_{\ell_j}C_{\ell_j}\}, \tilde{\mathcal{G}}=\{\tilde{g}_{\alpha j}\},\ F=\{F_\alpha\}\ ,
\ee
\be
\mathcal{M}=G\mathcal{G}K\tilde{\mathcal{G}}^TF,
\ee
Eq.\ \eqref{ampr} is equivalent to
\be
\label{es}
[1\!\mathrm{I}-(\lambda/r_0)^2\mathcal{M}]A=0\ ,
\ee
and poles may be found by computing the determinant
\be
\label{det}
\det[1\!\mathrm{I}-(\lambda/r_0)^2\mathcal{M}]=0\ .
\ee
{Equation \eqref{det} is obtained by setting the boundary conditions for a set of eigenvalue equations and it generates real eigenvalues for $E$ or $k$ when the $MM$ channels are closed. When some $MM$ threshold opens, it analytically continues from a real region for $E$ to a complex area, and the imaginary part is interpreted as the decay width.} 
For the particular case where $n=2$, i.e., $\alpha=1,2$, the pole condition \eqref{det} is
\be
\small
\label{polesg}
\begin{split}
&\bigg[1-\bigg(\frac{\lambda}{r_0}\bigg)^2G_1F_1\sum_j\mu_jJ_{\ell j}C_{\ell j}g_{1j}\tilde{g}_{1j}\bigg]\bigg[1-\bigg(\frac{\lambda}{r_0}\bigg)^2G_2F_2\sum_j\mu_jJ_{\ell j}C_{\ell j}g_{2j}\tilde{g}_{2j}\bigg]\\
&=\bigg[\bigg(\frac{\lambda}{r_0}\bigg)^2G_2F_1\sum_j\mu_jJ_{\ell j}C_{\ell j}g_{2j}\tilde{g}_{1j} \bigg]\bigg[\bigg(\frac{\lambda}{r_0}\bigg)^2G_1F_2\sum_j\mu_jJ_{\ell j}C_{\ell j}g_{1j}\tilde{g}_{2j}\bigg]\ ,\\
\end{split}
\ee
with all functions defined at the point $r_0$. For the amplitudes, we divide Eq.\ \eqref{ampr} and $\tilde{A}_j$ in Eq.\ \eqref{ampg} by $A_1$, defined as $A_\alpha$ in Eq.\ \eqref{ampr}. We get
\be
\label{ampsg}
\begin{split}
&A_\alpha=\frac{G_\alpha(r_0)}{G_1(r_0)}\frac{\sum_j g_{\alpha j}\mu_j J_j(kr_0)C_j(kr_0)\sum_\beta\tilde{g}_{j\beta}F_\beta(r_0)A_\beta}{\sum_j g_{1j}\mu_jJ_j(kr_0)C_j(kr_0)\sum_\beta\tilde{g}_{j\beta}F_\beta(r_0)A_\beta}A_1\ ,\\[2mm]
&A_j=\frac{r_0}{\lambda G_1(r_0)}\frac{\mu_jC_j(kr_0)\sum_\beta \tilde{g}_{j\beta} F_\beta(r_0)A_\beta}{\sum_j g_{1j}\mu_jJ_j(kr_0)C_j(kr_0)\sum_\beta\tilde{g}_{1\beta}F_\beta(r_0)A_\beta}A_1\ .
\end{split}
\ee
The amplitudes in Eq.\ \eqref{ampsg}, along with the set
\be
\label{amp4}
\begin{split}
&A_1, \ \ B_\alpha=\frac{F_\alpha(r_0)}{G_\alpha(r_0)}A_\alpha\\
&B_j=\frac{J_j(kr_0)}{C_j(kr_0)}A_j\ ,
\end{split}
\ee
are completely determined for $n=2$, in function of $A_1$. Finally, the normalization constant $\mathcal{N}$ of the total wave-function is determined by 
\be
\label{n}
\int_0^\infty \!dr\ |u(r)|^2=
\sum_\alpha\int_0^\infty \!dr\ |u_\alpha(r)|^2+\sum_j\int_0^\infty \!dr\ |u_j(r)|^2=\mathcal{N}^2 \; .
\ee



\begin{thebibliography}{99}


\bibitem{prl108p122001}
  M.~Bondar {\it et al.} (Belle Collaboration),
  \href{http://journals.aps.org/prl/abstract/10.1103/PhysRevLett.108.122001}{Phys.\ Rev.\ Lett.\ {\bf 108}, 122001 (2012)}; 
  P.~Krokovny {\it et al.} (Belle Collaboration),
  \href{http://journals.aps.org/prd/abstract/10.1103/PhysRevD.88.052016}{Phys.\ Rev.\ D {\bf 88}, 052016 (2013)}.  

\bibitem{prl110p252001}
  M.~Ablikim {\it et al.} (BESIII Collaboration),
  \href{http://journals.aps.org/prl/abstract/10.1103/PhysRevLett.110.252001}{Phys.\ Rev.\ Lett.\  {\bf 110}, 252001 (2013)}; 
  \href{http://journals.aps.org/prl/abstract/10.1103/PhysRevLett.115.112003}{Phys.\ Rev.\ Lett.\ {\bf115}, 112003 (2015)}; 
  Z.Q.~Liu {\it et al.} (Belle Collaboration),
  \href{http://journals.aps.org/prl/abstract/10.1103/PhysRevLett.110.252002}{Phys.\ Rev.\ Lett.\  {\bf 110}, 252002 (2013)}.
  
\bibitem{prl112p022001}
  M.~Ablikim {\it et al.} (BESIII Collaboration),
  \href{http://journals.aps.org/prl/abstract/10.1103/PhysRevLett.112.022001}{Phys.\ Rev.\ Lett.\ {\bf 112}, 022001 (2014)};  
  \href{http://journals.aps.org/prl/abstract/10.1103/PhysRevLett.115.222002}{Phys.\ Rev.\ Lett.\ {\bf 115}, 222002 (2015)}. 
   
\bibitem{prl111p242001}
  M.~Ablikim {\it et al.} (BESIII Collaboration),
  \href{http://journals.aps.org/prl/abstract/10.1103/PhysRevLett.111.242001}{Phys.\ Rev.\ Lett.\ {\bf 111}, 242001 (2013)};
  \href{http://journals.aps.org/prl/abstract/10.1103/PhysRevLett.112.132001}{Phys.\ Rev.\ Lett.\ {\bf 112}, 132001 (2014)};  \href{http://journals.aps.org/prl/abstract/10.1103/PhysRevLett.115.182002}{Phys.\ Rev.\ Lett.\ {\bf 115}, 182002 (2015)};
  \href{http://journals.aps.org/prl/abstract/10.1103/PhysRevLett.113.212002}{Phys.\ Rev.\ Lett.\ {\bf 113}, 212002 (2014)}.  

\bibitem{pdg}
  K.~A.~Olive {\it et al.} (Particle Data Group),
  \href{}{Chin.\ Phys.\ C {\bf 38}, 090001 (2014).}

\bibitem{prd78p072004}
  R.~Mizuk {\it et al.} (Belle Collaboration),
  \href{http://journals.aps.org/prd/abstract/10.1103/PhysRevD.78.072004}{Phys.\ Rev.\ D {\bf 78}, 072004 (2008)}.  


\bibitem{Zbtetra}
C.Y.\ Cui, Y.L.\ Liu, and M.Q.\ Huang,
\href{http://journals.aps.org/prd/abstract/10.1103/PhysRevD.85.074014}{Phys.\ Rev.\ D {\bf 85}, 074014 (2012)};
A.\ Ali, C.\ Hambrock, and W.\ Wang,
\href{http://journals.aps.org/prd/abstract/10.1103/PhysRevD.85.054011}{Phys.\ Rev.\ D {\bf85}, 054011 (2012)}.

\bibitem{Zctetra}
C.R.\ Deng, J.L.\ Ping, H.X.\ Huang, and F.\ Wang,
\href{http://journals.aps.org/prd/abstract/10.1103/PhysRevD.90.054009}{Phys.\ Rev.\ D {\bf 90}, 054009 (2014)}; 
Z.G.\ Wang and T. Huang,
\href{http://journals.aps.org/prd/abstract/10.1103/PhysRevD.89.054019}{Phys.\ Rev.\ D {\bf 89}, 054019 (2014)};
J.M.\ Dias, F.S.\ Navarra, M.\ Nielsen, and C.M.\ Zanetti,
\href{http://journals.aps.org/prd/abstract/10.1103/PhysRevD.88.016004}{Phys.\ Rev.\ D {\bf 88}, 016004 (2013)};
Zhi-Gang Wang,
\href{http://iopscience.iop.org/article/10.1088/0253-6102/63/3/325/meta}{Commun.\ Theor.\ Phys.\ {\bf63}, 325 (2015)}.

\bibitem{prl111p162003}
Eric Braaten,
\href{http://journals.aps.org/prl/abstract/10.1103/PhysRevLett.111.162003}{Phys.\ Rev.\ Lett.\ {\bf 111}, 162003 (2013)}.

\bibitem{prd90p016003}
F.~Aceti, M.\ Bayar, E.\ Oset, A.M.\ Torres, K.P.\ Khemchandani, J.M.\ Dias, F.S.\ Navarra, and M.\ Nielsen,
\href{http://journals.aps.org/prd/abstract/10.1103/PhysRevD.90.016003}{Phys.\ Rev.\ D {\bf 90}, 016003 (2014)}.

\bibitem{prd92p034004}
Jun He,
\href{http://journals.aps.org/prd/abstract/10.1103/PhysRevD.92.034004}{Phys.\ Rev.\ D {\bf 92}, 034004 (2015)}.

\bibitem{1511.06779}
R.A.\ Briceno et {\it al.},
\href{http://arxiv.org/abs/1511.06779}{arXiv:1511.06779 [hep-ph].}

\bibitem{ppnp67p449}
G.\ Rupp, S.\ Coito, and E.\ van Beveren,
\href{http://www.sciencedirect.com/science/article/pii/S0146641012000105}{Prog.\ Part.\ Nucl.\ Phys.\ {\bf 67}, 449 (2012).}

\bibitem{prd84p054002}
  Z.F.\ Sun, J.\ He, X.\ Liu, Z.G.~Luo, and S.L.~Zhu,
  \href{http://journals.aps.org/prd/abstract/10.1103/PhysRevD.84.054002}{Phys.\ Rev.\ D  {\bf 84}, 054002 (2011)};
  Y.C.~Yang, J.L.~Ping, C.R.~Deng, and H.S.~Zong,
  \href{http://iopscience.iop.org/article/10.1088/0954-3899/39/10/105001/meta}{{J.\ Phys.\ G} {\bf 39}, 105001 (2012)}.

\bibitem{prd90p076008}
  Jun He,
  \href{http://journals.aps.org/prd/abstract/10.1103/PhysRevD.90.076008}{Phys.\ Rev.\ D {\bf 90}, 076008 (2014)};
  H.W.\ Ke, X.Q.\ Li, Y.L.\ Shi, G.L.\ Wang, and X.H.\ Yuan,
  \href{http://link.springer.com/article/10.1007%2FJHEP04%282012%29056}{JHEP {\bf 04}, 056 (2012)}.

\bibitem{epjc73p2561}
H.W.\ Ke, Z.T.\ Wei, X.Q.\ Li,
\href{http://epjc.epj.org/articles/epjc/abs/2013/10/10052_2013_Article_2561/10052_2013_Article_2561.html}{Eur.\ Phys.\ J.\ C {\bf 73}, 2561 (2013)}.

  
\bibitem{prd89p094506}
 Ying Chen {\it et al.} (CLQCD Collab.),
 \href{http://journals.aps.org/prd/abstract/10.1103/PhysRevD.89.094506}{Phys.\ Rev.\ D {\bf 89}, 094506 (2014)}.

\bibitem{plb727p172}
S.\ Prelovsek, and L.\ Leskovec,
\href{http://www.sciencedirect.com/science/article/pii/S0370269313008022}{Phys.\ Let.\ B {\bf 727}, 172 (2013)}; S.\ Prelovsek, C.B.\ Lang, L.\ Leskovec, and D.\ Mohler, \href{http://journals.aps.org/prd/abstract/10.1103/PhysRevD.91.014504}{Phys.\ Rev.\ D {\bf 91}, 014504 (2015)}.

\bibitem{1411.1389}
S.H.\ Lee, C.\ DeTar, D.\ Mohler, H. Na, [Fermilab Lattice and MILC Collaborations],
\href{http://arxiv.org/abs/1411.1389}
 {{arXiv:1411.1389 [hep-lat]}}. 

\bibitem{molsrules}
Jian-Rong Zhang,
\href{http://journals.aps.org/prd/abstract/10.1103/PhysRevD.87.116004}{Phys.\ Rev.\ D {\bf 87}, 116004 (2013)};
C.Y.~Cui, X.H.~Liao, Y.L.~Liu, and M.Q.~Huang, 
\href{http://iopscience.iop.org/article/10.1088/0954-3899/41/7/075003/meta}{J.\ Phys.\ G {\bf 41}, 075003 (2014)};
Z.G.\ Wang and T.\ Huang,
\href{http://epjc.epj.org/articles/epjc/abs/2014/05/10052_2014_Article_2891/10052_2014_Article_2891.html}{Eur.\ Phys.\ J.\ C {\bf 74}, 2891 (2014)};
W.\ Chen, T.G.\ Steele, H.X.\ Chen, and S.L.\ Zhu,
\href{http://journals.aps.org/prd/abstract/10.1103/PhysRevD.92.054002}{Phys.\ Rev.\ D {\bf 92}, 054002 (2015)}.

\bibitem{epl96p11002}
  David Bugg, 
  \href{http://iopscience.iop.org/article/10.1209/0295-5075/96/11002/meta;jsessionid=DB2311CF2FACB85A977E46B6810726D9.c3.iopscience.cld.iop.org}{Eur.\ Phys.\ Lett.\ {\bf 96}, 11002 (2011)}.
  
\bibitem{prd91p034009}
  Eric S.~Swanson,
  \href{http://journals.aps.org/prd/abstract/10.1103/PhysRevD.91.034009}{Phys.\ Rev.\ D {\bf 91}, 034009 (2015)}.

\bibitem{plb747p410}
  Adam P.\ Szczepaniak, 
  \href{http://www.sciencedirect.com/science/article/pii/S0370269315004517}{Phys.\ Lett.\ B {\bf 747}, 410 (2015)}.

\bibitem{plb753p297}
  X.H.\ Liu, M.\ Oka, and Q.\ Zhao, 
  \href{http://www.sciencedirect.com/science/article/pii/S0370269315004517}{Phys.\ Lett.\ B {\bf 753}, 297 (2016)}.

\bibitem{prd84p094003}
  D.Y.\ Chen and X.\ Liu,
  \href{http://journals.aps.org/prd/abstract/10.1103/PhysRevD.84.094003}{Phys.\ Rev.\ D  {\bf 84}, 094003 (2011)}; 
  D.Y.\ Chen, X.\ Liu, and T.\ Matsuki, 
  \href{http://journals.aps.org/prd/abstract/10.1103/PhysRevD.88.036008}{Phys.\ Rev.\ D  {\bf 88}, 036008 (2013)}.

\bibitem{prd91p051504}
  F.K.\ Guo, C.\ Hanhart, Q.\ Wang, and Q.\ Zhao,
  \href{http://journals.aps.org/prd/abstract/10.1103/PhysRevD.91.051504}{Phys.\ Rev.\ D  {\bf 91}, 051504 (2015)}; 
  M.\ Albaladejo, F.K.\ Guo, C.\ Hidalgo-Duque, and J.\ Nieves,
  \href{http://arxiv.org/abs/1512.03638}{arXiv: 1512.03638 [hep-ph]}.

\bibitem{prl111p132003}
Q.\ Wang, C.\ Hanhart, and Q.\ Zhao,
\href{http://journals.aps.org/prl/abstract/10.1103/PhysRevLett.111.132003}{Phys.\ Rev.\ Lett.\ {\bf 111}, 132003 (2013)};
X.H.\ Liu and G.\ Li,
\href{http://journals.aps.org/prd/abstract/10.1103/PhysRevD.88.014013}{Phys.\ Rev.\ D {\bf 88}, 014013 (2013)}.

\bibitem{epjc73p2351}
S.~Coito, G.~Rupp, and E.~van Beveren, 
\href{http://epjc.epj.org/articles/epjc/abs/2013/03/10052_2013_Article_2351/10052_2013_Article_2351.html}{Eur.\ Phys.\ J.\ C  {\bf 73}, 2351 (2013)}.

\bibitem{epjc75p26}
M.~Cardoso, G.~Rupp, and E.~van Bevern,
\href{http://epjc.epj.org/articles/epjc/abs/2015/01/10052_2014_Article_3254/10052_2014_Article_3254.html}{Eur.\ Phys.\ J.\ C  {\bf 75}, 26 (2015)}.

\bibitem{zpc19p275}
E.~van Beveren, C.\ Dullemond, and  T.A.\ Rijken  
\href{http://link.springer.com/article/10.1007%2FBF01572256}{Z.\ Phys.\ C {\bf 19}, 275 (1983)}.

\bibitem{prd27p1527}
E.~van Beveren, G.\ Rupp, T.A.\ Rijken, and C.\ Dullemond,  
\href{http://journals.aps.org/prd/abstract/10.1103/PhysRevD.27.1527}{Phys.\ Rev.\ D {\bf 27}, 1527 (1983)}.

\bibitem{apppsb5p1007}
G.\ Rupp, S.\ Coito, E.\ van Beveren,
\href{http://www.actaphys.uj.edu.pl/fulltext?series=Sup&vol=5&page=1007}{Act.\ Phys.\ Pol.\ Proc.\ Supp.\ B {\bf 5}, 1007 (2012)}.

\bibitem{prl91p012003}
E.\ van Beveren and G.\ Rupp,
\href{http://journals.aps.org/prl/abstract/10.1103/PhysRevLett.91.012003}{Phys. Rev. Lett. {\bf 91}, 012003 (2003)};
\href{http://journals.aps.org/prl/abstract/10.1103/PhysRevLett.97.202001}{Phys. Rev. Lett. {\bf 97}, 202001 (2006)};
S.~Coito, G.~Rupp, and E.~van Beveren, 
\href{http://journals.aps.org/prd/abstract/10.1103/PhysRevD.84.094020}{Phys.\ Rev.\ D  {\bf 84}, 094020 (2011)}.

\bibitem{pcoup}
Eef van Beveren,  
\href{http://link.springer.com/article/10.1007%2FBF01574179}{Z.\ Phys.\ C {\bf 17}},\href{http://link.springer.com/article/10.1007%2FBF01577044}{ 135 (1983); Z.\ Phys.\ C {\bf 21}, 291 (1984)}.

\bibitem{posp003}
E.~van Beveren and G.~Rupp,  
\href{http://pos.sissa.it/archive/conferences/128/003/HQL%202010_003.pdf}{PoS HQL2010, 003 (2010)}.

\bibitem{plb669p156}
T.\ Matsuki, T.\ Morii, and K.\ Sudoh,
\href{http://www.sciencedirect.com/science/article/pii/S0370269308011842?np=y}{Phys.\ Let. B {\bf669}, 156 (2008)}.

\bibitem{prd84p054010}
A.E.\ Bondar, A.\ Garmash, A.I.\ Milstein, R.\ Mizuk, and M.B.\ Voloshin,
\href{http://journals.aps.org/prd/abstract/10.1103/PhysRevD.84.054010}{Phys.\ Rev.\ D {\bf 84}, 054010 (2011)}.

\end{thebibliography}
\end{document}